\documentclass[sigplan,screen,nonacm]{acmart}
\usepackage{soul}
\usepackage{listings}
\usepackage{natbib}

\lstdefinestyle{outputStyle}{
    belowcaptionskip=1\baselineskip,
    breaklines=true,
    frame=none,
    numbers=none,
    basicstyle=\footnotesize\ttfamily,
    keywordstyle=\bfseries\color{green!40!black},
    commentstyle=\itshape\color{purple!40!black},
    identifierstyle=\color{blue},
    backgroundcolor=\color{gray!10!white},
}

\AtBeginDocument{%
  \providecommand\BibTeX{{%
    \normalfont B\kern-0.5em{\scshape i\kern-0.25em b}\kern-0.8em\TeX}}}

\setcopyright{acmcopyright}
\copyrightyear{2024}
\acmYear{2024}

\begin{document}

\newcommand\sudip[1]{\textcolor{blue}{Sudip : #1}}
\newcommand\ivan[1]{\textcolor{orange}{Ivan : #1}}
\newcommand\martin[1]{\textcolor{purple}{Martin : #1}}
\title{Utilizing Large Language Models to Translate RFC Protocol Specifications to CPSA Definitions}


\author{Martin Duclos}
\affiliation{%
 \institution{Mississippi State University}
  \country{}
}\email{md128@msstate.edu}

\author{Ivan A. Fernandez}
\affiliation{%
 \institution{Mississippi State University}
\country{}
}\email{iaf28@msstate.edu}

 \author{Kaneesha Moore}
\affiliation{%
 \institution{Mississippi State University}
  \country{}
}\email{kkm267@msstate.edu}

 \author{Sudip Mittal}
\affiliation{%
 \institution{Mississippi State University}
  \country{}
}\email{mittal@cse.msstate.edu}

 \author{Edward Zieglar}
\affiliation{%
 \institution{National Security Agency}
   \country{}
}\email{evziegl@uwe.nsa.gov}

\renewcommand{\shortauthors}{Duclos et al.}


\begin{abstract}
This paper proposes the use of Large Language Models (LLMs) for translating Request for Comments (RFC) protocol specifications into a format compatible with the Cryptographic Protocol Shapes Analyzer (CPSA). This novel approach aims to reduce the complexities and efforts involved in protocol analysis, by offering an automated method for translating protocol specifications into structured models suitable for CPSA. In this paper we discuss the implementation of an RFC Protocol Translator, its impact on enhancing the accessibility of formal methods analysis, and its potential for improving the security of internet protocols.
\end{abstract}
\maketitle

\section{Introduction \& Background}
The current standards process by which the Internet Engineering Task Force (IETF) reviews security protocols is hindered by a lack of verification of stated properties. This shortcoming is caused by the absence of formal proofs, crucial for proper verification. In its effort to improve the standards process, the IETF is encouraging protocol developers to incorporate formal analysis and validation into their work \cite{ietf2023ufmrg}. However, formal analysis is complex and requires significant expertise, rendering it inaccessible to some protocol developers \cite{gritzalis1999security}. This paper aims to support the IETF's efforts by simplifying the tasks of formal analysis and validation, making them more accessible to protocol developers. The contributions of this research project includes the creation of tools for integration of formal analysis into a developer's workflow, enhancing the security of future protocols.

One approach to formal analysis, described by Meadows \cite{meadows1992applying} and originally suggested by Kemmerer \cite{kemmerer1989using}, involves modeling a protocol in a formal language. A verification system is then used to validate models against the stated protocol properties. Protocol developers can perform protocol analysis and verification using tools like Cryptographic Protocol Shapes Analyzer (CPSA), Maude-NPA, Tamarin, and ProVerif \textit{offline}. This task is both tedious and labor intensive, necessitating a \textit{domain expert} to parse and convert protocol specifications into models compatible with a formal methods tool \cite{ietf2023ufmrg}. As such, it is difficult to accurately convert protocol specifications into a form usable with a formal methods tool \cite{gritzalis1999security}.

The current IETF process could be improved by making the tools necessary for formal analysis more accessible. This would encourage protocol developers to integrate these tools into their workflow. In turn, this would make it easier for developers to perform protocol analysis while developing specifications and prior to submitting a Request for Comments (RFC). An improved workflow could include an `RFC proposal translator system' that could accept a set of protocol specifications and `translate' them into formal methods tool input candidates: a `protocol definition' that can then be processed and the resulting output displayed alongside the specifications. Such a system would automate the generation of protocol definitions compatible with formal methods tools. This RFC proposal translator is designed to assist with one of the goals of the Usable Formal Methods Research Group (UFMRG) established by the IETF in January 2023 \cite{ietf2023ufmrg}, which is to understand how formal methods can be incorporated into the development of specifications for security protocols.

In this paper, we propose an RFC proposal translator that leverages the recent advances in Natural Language Processing (NLP) and Generative AI (GenAI). Our proposed system inputs specifications into a code-specialized Large Language Model (LLM), which then automatically generates candidate structured protocol definitions. This system is designed with CPSA as the target protocol analyzer, necessitating the LLM to output in CPSA syntax. The CPSA formal methods tool requires a protocol (protocol definition) and a partial description of an execution (protocol skeleton) as input, in the form of structured text using symbolic expressions (s-expressions) \cite{cpsa41manual2023}. The structured output candidates from the LLM will need to be revised by a domain expert before being processed by CPSA, ensuring an accurate representation of the protocol specifications. This novel approach will improve the overall efficiency of the current process by reducing manual effort and complexity associated with translating the specifications to the format required by CPSA, consequently making formal analysis more accessible.

\section{RFC Proposal
Translator Architecture}
This section describes the system architecture of our RFC Proposal Translator that leverages a Code-Specialized LLM (CSLLM) (See Figure \ref{fig:system_architecture}). The main objective of the CSLLM is to aid with the translation of English-written protocol specifications into structured text made of symbolic expressions. This structured text can subsequently serve as input for CPSA. System components include:

\begin{figure}[ht]
    \centering
    \includegraphics[width=1\linewidth]{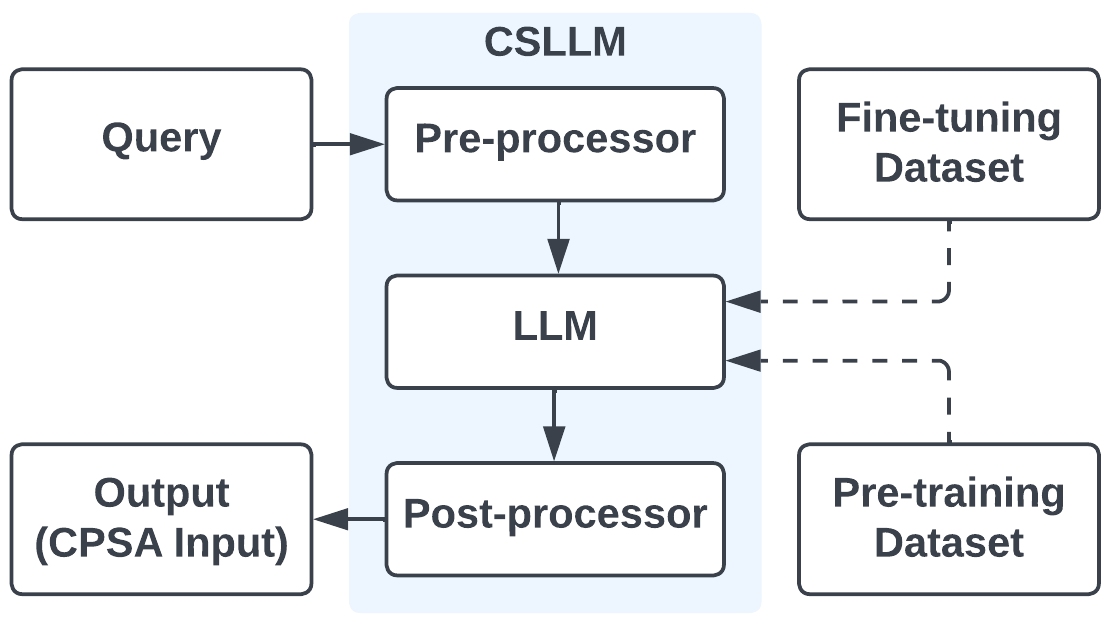}
    \caption{RFC Proposal
Translator Architecture}
\label{fig:system_architecture}
\vspace{-3mm}
\end{figure}

    \noindent\textbf{1) Query:} A user-provided query serves as both an input and a source of knowledge for the LLM. 
    
    \noindent\textbf{2) Pre-processor:} This component verifies that a query forwarded to the LLM meets specific conditions for processing.
    
    \noindent \textbf{3) LLM:} The LLM component used for this work, consists of Meta's Code Llama \cite{roziere2023code}. The LLM is responsible for transforming the input query into an output that can be processed by CPSA. For this purpose, the LLM relies on three types of knowledge: 1) knowledge acquired during the initial training phase, 2) knowledge gained through the fine-tuning dataset, and 3) context-specific knowledge provided by the user as part of the query and pre-processing phase.
    
    \noindent \textbf{4) Pre-training Dataset:} The initial dataset used to train Code Llama consists of 500 billion tokens, composed of publicly available code, discussions about code, and code snippets included in natural language questions or answers \cite{roziere2023code}.
    
    \noindent \textbf{5) Fine-tuning Dataset:} The fine-tuning dataset comprises three types of data: 1) a curated set of RFCs, 2) CPSA protocol definitions, and 3) matching pairs of protocol requirements and CPSA protocol definitions. Code Llama uses these customized datasets to gain an understanding of RFCs, message algebra and s-expressions.
    
    \noindent \textbf{6) Post-Processor:} This component ensures the output generated by Code Llama complies with conditions to minimize the risk of CPSA rejecting the output for syntactic reasons.
    
    \noindent \textbf{7) Output:} The output generated by Code Llama consists of structured text composed of s-expressions which form a CPSA protocol definition and corresponding skeletons.

\section{Experimental Results}
The results below were obtained using the 34B parameters version of Code Llama: We submitted the following query to the pre-trained LLM:\emph{"Using Diffie-Hellman algebra and s-expressions, define a CPSA-compatible input file for a protocol to allow Alice and Bob to exchange a meeting location."} A partial output representing Alice's point of view from this first step can be seen in Listing \ref{lst:pretraining}. Subsequently, we injected the content of two CPSA definition files in the context window of the LLM and re-submitted the query above. A partial output representing Alice's point of view of this second step can be seen in Listing \ref{lst:finetuning}. We can observe that Listing \ref{lst:pretraining} fails to generate an output resembling a CPSA input file. In contrast, injecting two CPSA definition files enabled the LLM to generate an output (See Listing \ref{lst:finetuning}) containing s-expressions and Diffie-Hellman algebra.

\begin{lstlisting}[caption=Partial LLM Output Utilizing Pre-training Dataset, label={lst:pretraining}, style=outputStyle,basicstyle=\scriptsize]
# Alice's actions
Alice:
 - Generate a public/private key pair
 - Send her public key to Bob
 - Generate a shared secret key using Bob's public key
 - Encrypt the meeting location using shared secret key
 - Send the encrypted meeting location to Bob
\end{lstlisting}

\begin{lstlisting}[caption=Partial LLM Output Utilizing Fine-tuning Dataset, label={lst:finetuning}, style=outputStyle,basicstyle=\scriptsize]
(defprotocol meeting-location
 (defrole alice
    (vars (x y expn) (b name) (n text) (kh akey))
    (trace
     (send (enc (exp (gen) x) (pubk b)))
     (recv (cat (exp (gen) y) (hash (cat (exp (exp (gen) x) y)))))
     (send (enc n (exp (exp (gen) y) x))))
    (pen-non-orig x)
    (uniq-orig n))
\end{lstlisting}

\section{Conclusion}
In conclusion, this paper contributes to the field of protocol analysis by using LLMs to translate RFC specifications into a CPSA compatible format, thereby improving the overall accessibility of formal methods analysis.

\bibliographystyle{unsrt}
\bibliography{bibliography}
\end{document}